\documentclass[aps,prb,twocolumn,superscriptaddress]{revtex4}
\usepackage{natbib}
\usepackage{graphicx}
\usepackage{amssymb}

\usepackage{amsmath}

\begin{document}

\title{Microwave frequency modulation in continuous-wave far-infrared ESR utilizing a quasioptical reflection bridge}

\author{B\'alint N\'afr\'adi}
\email[]{balint.nafradi@yahoo.com}
\affiliation{Institute of Physics of Complex Matter, FBS, Swiss Federal Institute of Technology (EPFL), CH-1015 Lausanne, Switzerland}
\author{Richard Ga\'al}
\affiliation{Institute of Physics of Complex Matter, FBS, Swiss Federal Institute of Technology (EPFL), CH-1015 Lausanne, Switzerland}
\author{Titusz Feh\'er}
\affiliation{Institute of Physics of Complex Matter, FBS, Swiss Federal Institute of Technology (EPFL), CH-1015 Lausanne, Switzerland}
\affiliation{Institute of Physics, Budapest University of Technology and Economics, and Condensed Matter Research Group of the Hungarian Academy of Sciences, P.O.Box 91, H-1521 Budapest, Hungary}
\author{L\'aszl\'o Forr\'o}
\affiliation{Institute of Physics of Complex Matter, FBS, Swiss Federal Institute of Technology (EPFL), CH-1015 Lausanne, Switzerland}

\date{\today}

\begin{abstract}
We report the development of the frequency-modulation (FM) method for measuring electron spin resonance (ESR) absorption in the 210-420 GHz frequency range.
We demonstrate that using a high-frequency ESR spectrometer without resonating microwave components enables us to overcome technical difficulties associated with the FM method due to nonlinear microwave-elements, without sacrificing spectrometer performance.
FM was achieved by modulating the reference oscillator of a 13~GHz Phase Locked Dielectric Resonator Oscillator, and amplifying and frequency-multiplying the resulting millimeter-wave radiation up to 210, 315 and 420 GHz.
ESR spectra were obtained in reflection mode by a lock-in detection at the fundamental modulation frequency, and also at the second and third harmonic.
Sensitivity of the setup was verified by conduction electron spin resonance measurement in KC$ _{60}$.
\end{abstract}

\maketitle

\section{Introduction}
\label{intro}
In the vast majority of continuous-wave (CW) electron spin resonance (ESR) spectrometers today, magnetic field modulation and synchronous harmonic detection is used to achieve high sensitivity.
A small amplitude magnetic field is applied parallel to the DC field and based on the ``transfer of modulation" principle the fundamental frequency and harmonics of the magnetic field modulation appears on the envelope of the microwave carrier only during magnetic resonance.
Stability of the spectrometer is based in large part on satisfying this principle.
The time-varying magnetic field generates a certain number of problems: eddy currents in the components made of conductive materials can cause substantial heating; even more serious problem is the torque arising from the interaction of the three magnetic fields (DC, modulation and that of the eddy currents) which may cause vibrations and increase noise through microphonics.
This problem is accentuated in high-field ESRs because the DC field is over an order of magnitude higher than that used in conventional X-band spectrometers.

There are various methods for detecting ESR signals without magnetic field modulation: pulsed ESR spectroscopy, multiquantum ESR spectroscopy, longitudinally detected ESR spectroscopy, thermal modulation, and amplitude modulation of the microwave source.
These are advanced methods for ESR instrumentation that require specific microwave technologies for the spectrometer setup and raise technical issues in ESR applications.

On the other hand, sinusoidal modulation of the microwave frequency that is incident on the resonator is widely used in automatic frequency control (AFC) systems as part of a feedback
circuit to lock the oscillator to the resonance frequency of the resonator. 
In this application of frequency modulation (FM), the frequency modulation depth, $ \delta\omega_m $, is low. When $ \delta\omega_m $ becomes high, FM followed by microwave detection and subsequent
lock-in detection can be used to obtain an ESR spectrum.
The FM depth is related to the equivalent field modulation amplitude, $ B_m $, by Eq. [\ref{eq:1}]:
    \begin{equation}
        \label{eq:1}
        \delta \omega_m = \gamma B_m
    \end{equation}
where $ \gamma $ is the gyromagnetic ratio of the electron.

A recent paper by Kalin et al. \cite{Kalin2003} provides theoretical arguments which support the nonequivalence of magnetic field modulation and microwave FM methods.
These theoretical predictions were tested in Hirata et al. \cite{Hirata2004}.
They have found that the ESR spectra obtained by the FM method differ from those obtained by conventional field modulation method only when the modulation frequency is higher than the frequency corresponding to the linewidth. 
Under our experimental conditions, we conclude that frequency and field modulation are equivalent.

A high microwave frequency and low-Q resonator are favorable for FM experiments because the sample is irradiated with a nearly constant value of $ B_{1} $ over the range of frequency modulation depth.
An obstacle in the application of the FM method to low frequency ESR is the fact that those spectrometers have high quality factor  resonators to increase sensitivity, while in the case of FM a low-Q resonator should be used, otherwise the frequency response of the resonator induces a significant signal not related to the ESR.
Hirata et al. \cite{Hirata2002,Hirata2003} overcome this difficulty by using an electronically tunable L-band surface-coil type resonator with an unloaded Q-value of 260 and an automatic tuning control (ATC) and automatic matching control (AMC).
Hyde et al. \cite{Hyde2007} on the other hand used a W-band low-Q loop-gap type resonator with loaded Q-value of 90.

Here we report initial ESR results obtained by FM in the far-infrared frequency range.
The intrinsically higher sensitivity of high-field ESR allows us to dispense with microwave resonators altogether, and still have sufficient signal-to noise ratio: by this we overcome almost all of the technical difficulties that were reported previously in low-frequency FM measurements \cite{Hirata2002,Hirata2003,Hyde2007}.
\section{Experimental}
\label{exp}
    \begin{figure}
        \begin{center}
            \includegraphics*[width=7cm]{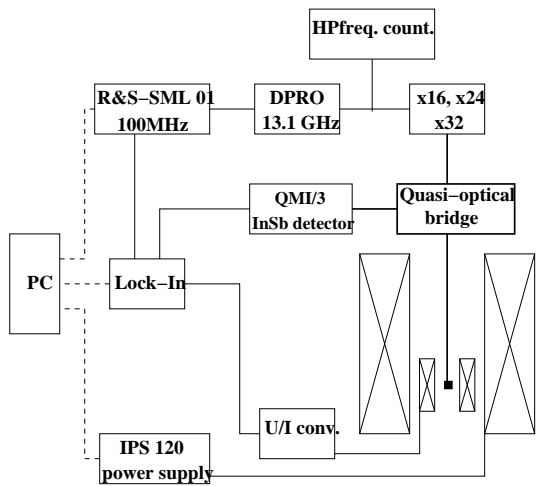}
        \end{center}
        \caption{Block diagram of the 210-420~GHz continuous-wave ESR spectrometer utilizing either frequency or field modulation. Dashed line shows the GPIB communication. More detail is given in the text.}
        \label{fig:1}
    \end{figure}
Our spectrometer is depicted in fig. \ref{fig:1}.
The frequency modulation is achieved by modulating the 100~MHz reference oscillator (Rohde\&Schwarz SML01) of a phase locked dielectric resonator oscillator (Herley-CTI DPRO).
The resulting 13.1~GHz carrier-frequency frequency-modulated radiation is amplified and frequency-multiplied with a chain of frequency doublers (Virginia Diodes) up to 420~GHz.
The frequency and phase-noise of the oscillator is determined by the noise of the R\&S reference oscillator.

The light is guided to the sample through a quasi-optical bridge (Thomas Keating Instruments) and back to a liquid helium cooled InSb hot-electron bolometer (QMC Instruments). 
The light inside the cryostat propagates in a HE$_{11}$ corrugated tube (Thomas Keating Instruments). 
The system is optimized to 210~GHz operation, but it also operates at higher millimeter-wave frequencies, because the tubing is oversized. 
The overall loss in the corrugated tubing around 210~GHz is less than 1~dB, which increases up to $\sim$20~dB while doubling the millimeter-wave frequency. 
This increased loss and decreasing oscillator power altogether causes worst signal to noise ratio at higher frequencies as shown in fig.~\ref{fig:4}.

The optical-path difference in our setup between the reference-arm and the sample arm is $\sim$5~m, what is much longer than the wavelength of our millimeter-wave radiation ($\lambda\sim$1.5~mm at 210~GHz), but the off-resonance millimeter-wave signal-level can be reduced at the detector to the electronic noise ground at least for half an hour during either conventional magnetic field or frequency-modulation, what shows that frequency and phase noise in our setup is not important. In a large part this is due to the fact that the bridge we use works with polarized light, and only ESR activity changes the polarization of the millimeter waves. A strategically placed grid polarizer assures that only microwaves with transverse polarization reach the mixer, and that off-resonance the reflected signal is truly zero. The bridge is a homodyne bridge, thus the path difference of 5~m induces maximum phase jitter of $10^{-6}$.

For data acquisition we use a standard field-sweep method in the present FM experiments.
An Oxford Instruments superconducting magnet (maximum field 16~T, homogeneity $ 10^{-5} /cm^3$) provides the quasi-static magnetic field $ B_{0} $.

For the standard magnetic field-modulated measurement we use a small solenoid (max. 8~mT peak-to-peak modulation amplitude at 20~kHz modulation frequency).
The field-modulation is driven by a home-made voltage-controlled current source (U/I converter). 
This converter is normally driven by an internal oscillator of Stanford Research systems SR830 lock-in amplifier, which is also used for harmonic detection.
For the frequency modulation we use an internal oscillator of the R\&S local reference oscillator. 
Data acquisition is done via GPIB bus with a PC. 
Acquisition of the magnetic field and the lock-in signal is synchronized using a simple function generator. 

All experiments presented in this paper were carried out on KC$_{60}$ powder, filled into a 2~mm inner diameter quartz tube.
The sample was located at the bottom of a corrugated waveguide, which is inside a variable temperature insert (VTI), in the center of the magnet. 

\section{Results and discussion}
\label{res}
    \begin{figure}
        \begin{center}
            \includegraphics[width=7.5cm]{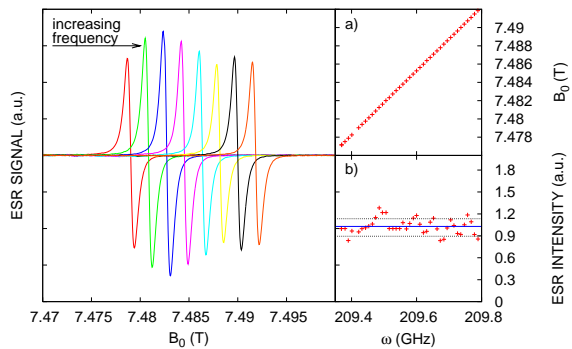}
        \end{center}
        \caption{Microwave frequency dependence of the field modulated spectra in the 209.42 to 209.78~GHz frequency range. Between the consecutive curves the frequency difference is 52.4~MHz (1.87~mT shift in resonance filed). a) The resonance field as a function of the microwave frequency ($\omega$). b) The integrated intensity as a function of the microwave frequency. Straight and dashed lines represent the mean value and the standard deviation around the mean, respectively.}
        \label{fig:2}
    \end{figure}
As a rule of thumb, the amplitude of the FM should be similar to the linewidth which in turn is a big proportion of the carrier frequency in low field ESR. 
This technical limitation can be removed at higher frequencies.
This is due to the fact that the ratio between the frequency modulation depth corresponding to the linewidth and the microwave carrier frequency becomes smaller at higher frequencies, and therefore attaining FM amplitude similar to the linewidth becomes easier.

Another advantage of using higher frequencies is that the response of the millimeter-wave components is more linear and causes less baseline problems.

In our experiment we have used the conduction electron spin resonance (CESR) signal of KC$_{60}$.
From the experimental standpoint, the polycrystalline KC$_{60}$ sample features a strong, nearly Lorentzian shaped CESR signal, which can readily be observed at room temperature.  
For this reason, this material has already been used for the calibration of several ESR systems
\cite{Sienkiewicz2005,Simon2005}.
It has 0.75~mT (21.02~MHz) peak-to-peak linewidth at 209.6~GHz.
This is about 0.01\% of the carrier microwave frequency.
Comparatively we are able to tune our microwave oscillator $ \pm 0.1 \% $ eitherside of the center frequency which is an order of magnitude wider.

As a demonstration of linearity of our quasi-optical setup we show several field-modulation measured spectra (fig. \ref{fig:2}), all recorded with the same spectrometer settings but with slightly different microwave frequencies.
The technical parameters were as follows: field-modulation frequency 20~kHz, 0.4~mT modulation amplitude, 2~mW microwave power, 30~ms time constant, 1000 points per scan, 30~s per scan, one single run.
The millimeter-wave frequency was changed from 209.42 to 209.78~GHz in 52~MHz steps.
The measured shift is linear within our experimental precision, the integrated intensity is constant with $ \pm $ 10\% systematic variation. 
This is most likely due to the positioning of the sample with respect to the reflecting bottom of our light guide.
This change in intensity can contribute to the baseline in FM measurement but the frequency range is two orders of magnitude broader than the frequency modulation depth needed to over-modulate the KC$_{60}$ line.

\begin{figure}
        \begin{center}
            \includegraphics[width=7.5cm]{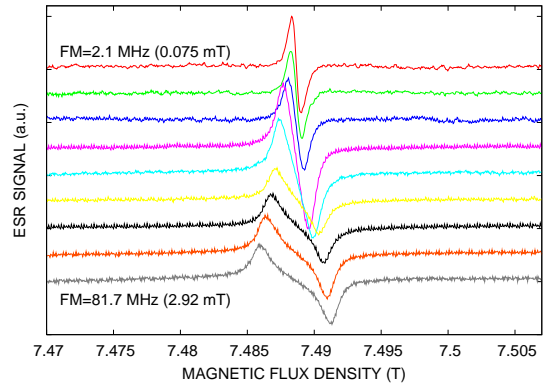}
        \end{center}
        \caption{Frequency-modulation depth dependence of the spectrum at 209.6~GHz carrier-frequency. The FM depth was increased from 2.2 to 81.74~MHz in 10~MHz steps. The spectra are shifted vertically for clarity.}
        \label{fig:3}
    \end{figure}

In fig. \ref{fig:3} we show a set of typical spectra recorded with frequency modulation amplitude up to eight times bigger than the intrinsic KC$_{60}$ linewidth.
During these measurements there was no significant baseline instability.
Technical parameters were the following: frequency-modulation frequency 1~kHz, 2~mW microwave power, 30~ms time constant, 1000 points per scan, 30~s per scan, one single run.
The modulation amplitude was changed from 2.2 to 81.744~MHz in 10~MHz steps, equivalent to 0.075--2.92~mT.
A qualitative change in the lineshape is visible between ESR spectra measured with 32~MHz and 52~MHz  modulation depth. This is the fingerprint of strong overmodulation of the KC$_{60}$ line.
At low modulation depth the signal to noise ratio with FM method is approximately ten times smaller than with field-modulation. It shows that FM method becomes more sensitive at high modulation depths what is unreachable with standard field modulation method ($>$~1~mT ).

\begin{figure}
        \begin{center}
            \includegraphics[width=7.5cm]{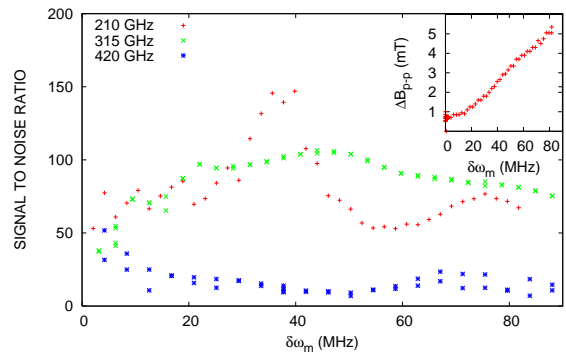}
        \end{center}
        \caption{Signal to noise ratio as a function of frequency modulation depth at 210, 310 and 420~GHz carrier frequency. The inset shows peak-to-peak linewidth as a function of frequency modulation depth at 210GHz.}
        \label{fig:4}
    \end{figure}

The measured modulation amplitude dependence of the peak-to-peak linewidth and signal-to-noise ratio (fig. \ref{fig:4}) followed the expected behavior for all the three carrier frequencies (210, 315 and 420 GHz).
The signal-to-noise ratio was calculated by dividing the amplitude of the signal by the standard deviation of the  baseline far away from the ESR line.


\begin{figure}
        \begin{center}
            \includegraphics[width=7.5cm]{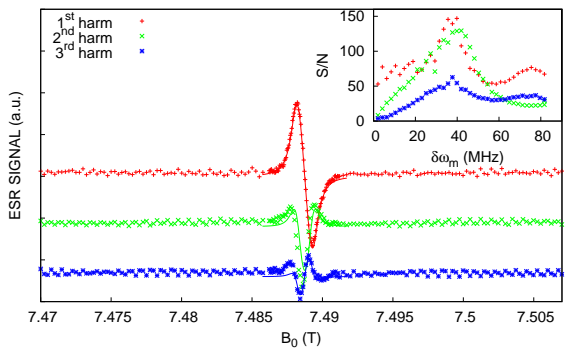}
        \end{center}
        \caption{Frequency-modulated spectra with different detected harmonics. The spectra are shifted vertically for clarity. The inset shows signal-to-noise ratio as a function of frequency modulation depth for all the three  detected harmonics. Lines are numerical simulations for higher harmonics detected signal.}
        \label{fig:5}
    \end{figure}

In addition to the first harmonic detected spectra, we measured frequency-modulated ESR signal on the second and third harmonics of the fundamental modulation frequency.
The frequency modulation depth during this measurement was 33.536~MHz (1.2~mT), all the other instrumental settings were identical to that of the data presented on fig.~\ref{fig:3}.
The resulting first, second and third derivative-like spectra are shown in fig.~\ref{fig:5}.
The expected suppression of the signal amplitude is clear, while no change in the noise was observed.
The inset in fig.~\ref{fig:5} shows signal-to-noise ratio as a function of frequency modulation depth in the case of all the detected harmonics.
\section{Conclusions}
\label{conc}
Our experimental findings suggest that the FM method at millimeter-wave frequencies is an alternative solution to achieve phase-sensitive detection when the side-effects of magnetic field modulation are detrimental to ESR detection.
These results show as well that stability can be achieved in frequency-modulated ESR at high frequencies because without resonating elements only the spins can change FM into AM when passing through the absorption.
The millimeter-wave frequency offers several advantages for frequency modulation CW ESR spectroscopy.
It is intrinsically broadband, permitting to forgo AMC and ATC circuits and allowing modulation frequencies at least as high as 100~MHz with p-p frequency modulation depth more than $ \sim $200~MHz.
Other benefits of the frequency modulation scheme are uniform modulation over the volume measured, freedom from microphonic noise and mechanical vibration due to the absence of magnetic field modulation.
The FM method can also solve the problem of heating in the modulation coils, and is therefore strongly applicable to high-pressure ESR measurements, where using modulation coils inside the pressure cell presents one of the biggest technical difficulties.
\section*{Acknowledgment}
\label{ackn}
This work was supported by the Swiss National Science Foundation and its NCCR MaNEP and the Hungarian state grants OTKA-PF63954 and K68807.
\end{document}